\documentclass[prl,preprintnumbers,amsmath,amssymb,twocolumn,superscriptaddress]{revtex4}

\usepackage{amsfonts}
\usepackage{amsmath}
\usepackage{amssymb}
\usepackage{color}
\usepackage{epsfig,bm,dcolumn}
\usepackage{framed}
\usepackage{float}
\usepackage{graphicx}
\usepackage{hyperref}
\usepackage{mathbbol}
\usepackage{mathrsfs}
\usepackage{multirow}
\usepackage{overpic}
\usepackage{slashed}
\usepackage{tabularx}
\usepackage[nottoc]{tocbibind}
\usepackage{varwidth}

\begin{document}
	\title{Neural Network Statistical Mechanics}
	\author{Lingxiao Wang}
	\affiliation{Department of Physics, Tsinghua University, Beijing 100084, China.}
	\affiliation{Frankfurt Institute for Advanced Studies, Ruth-Moufang-Str. 1, 60438 Frankfurt am Main, Germany}
	
	\author{Yin Jiang\footnote{jiang\_y@buaa.edu.cn}}
	\affiliation{Department of Physics, Beihang University, Beijing 100191, China.}

	\author{Kai Zhou\footnote{zhou@fias.uni-frankfurt.de}}
	\affiliation{Frankfurt Institute for Advanced Studies, Ruth-Moufang-Str. 1, 60438 Frankfurt am Main, Germany}
	\date{\today}

	\begin{abstract}
		We propose a general unsupervised framework to extract microscopic interactions from raw configurations with deep autoregressive neural networks. The approach constructs the modeling Hamiltonian by the neural networks, in which the interaction is encoded. The machine is trained with unlabeled data collected from \textit{Ab initio} computations or experiments. The well-trained neural networks reveal an accurate estimation to the possibility distribution of the configurations at fixed external parameters. It can be spontaneously extrapolated to detect the phase structures since classical statistical mechanics as prior knowledge here. We apply the approach to a 2-D spin system, training at a fixed temperature, and reproducing the phase structure. Scaling the configuration on lattice exhibits the interaction changes with the degree of freedom, which can be naturally applied to the experimental measurements. The framework bridges the gap between the real configurations and the microscopic dynamics with neural networks.
	\end{abstract}
	
	\maketitle
	

	{\it Introduction.---}
	In statistical thermodynamics, there are two main components are necessary for one to predict thermodynamic properties of a particular system, that is, the dependence of the micro-state distribution on the environment parameters, known as the Boltzmann factor $e^{-A/T}$ and the interaction details of the system, namely the Hamiltonian $A=H$, which is usually designed according to experimental and phenomenological properties of the given system combining with the intuition of theoretical physicists. Doubtlessly, the first one is more fundamental as a corollary of the principle of maximum entropy, which could be treated as one of the axioms of statistical physics. Besides the general dynamics, the second one, a specific model/Hamiltonian characterizing the system of interest, is computationally hard but indispensable\cite{cubitt:2012extracting}. Connecting the experimental data with models starts from a suitable choice of degree of freedom(\textit{d.o.f}). Usually it is chosen from either the experimental consideration or the aesthetic taste of physics, or both of them. Then motivated by the symbolic beauty and tractability, a concise model for the \textit{d.o.f} could have been developed and would be gradually decorated by taking more experimental facts into account, such as defects, boundary and different kinds of fluctuations\cite{anderson:2011more}.
	
	However in practice, the conciseness is not necessary if one could construct the interaction in an accurate and efficient enough way, such as with an elaborate neural network~\cite{carleo:2019machine,pfau:2020abinitio}, after all most of the analytically elegant models are not so innocent as they appear to be. And it is also natural to represent the complicated microscopic states by a generic machine, such as quantum simulators\cite{prufer:2020experimental} or intricate neural networks\cite{shen:2018selflearning}, where the key information of the state can be encoded efficiently, i.e., the wave-function \textit{Ansatz} was proposed with the corresponding 
	neural networks in solving quantum many-body problems~\cite{carleo:2017solving,yoshioka:2019constructing,hartmann:2019neuralnetwork,nagy:2019variational,sharir:2020deep,pfau:2020abinitio}. Furthermore, the generative models in machine learning were applied to generate the microscopic states\cite{urban:2018reducing,zhou:2019regressive,wang:2020recognizing}. These methods 
	bring improvements to the classical Metropolis Monte-Carlo algorithm~\cite{alexandru:2017deep, broecker:2017machine, mori:2018application,shen:2018selflearning}, such as the Generative Adversarial Networks(GANs) were applied to produce configurations on lattice~\cite{urban:2018reducing,zhou:2019regressive,singh:2020generative}. As some feasible alternatives, the Variational Auto-Encoder(VAE)~\cite{wetzel:2017unsupervised}, and the deep autoregressive networks~\cite{wu:2019solving,sharir:2020deep,wang:2020recognizing} show the reliable computing performance in both discrete and continuous systems, in which even the topological phase transition can be recognized\cite{wang:2020recognizing}. The neural networks with autoregressive property, such as masked neural networks and Pixel Convolutional Neural Networks(CNNs)~\cite{salimans:2017pixelcnn} or Recurrent Neural Networks(RNNs)~\cite{vandenoord:2016pixel} were embedded in the robust variational approach, which can reproduces the multifarious microscopic states effectively. Although above attempts have achieved meritorious improvements to generate micro-states, a further and attractive mission for neural networks, that is model-independently characterizing a general interaction purely basing on experimental data, is still uncompleted. For this ambitious and difficult problem,  there were some instructive researches, e.g. decoding the Sch\"ordinger equation from prepared configurations~\cite{wang:2019emergent}, extracting the many body interactions with the Restricted Boltzmann Machine(RBM)~\cite{rrapaj:2020exact}, and sampling equilibrium states by the Boltzmann generator based on flow model\cite{noe:2019boltzmann}. Besides, a research shown that the network can find the physically relevant parameters and exploit conservation laws to make predictions~\cite{iten:2020discovering}, which 
	is also close to our goals.

	In this letter we will explore a potential approach with the neural network to portray a generic interaction in statistical physics. There is no preset physical model, only classical statistical mechanics as a necessary prior knowledge\cite{hou:2020statistical}. First we will briefly review the ability of study the whole phase diagram for thermodynamics by making use of a large enough ensemble of microscopic configurations under a certain condition\cite{blickle:2007characterizing,li:2020hamiltonian}. This is also guaranteed by the Ergodic hypothesis. The bridge between the ensemble and each distribution of micro-states, and thus Hamiltonian, will be built with a special type of neural networks, the so-called autoregressive neural network. Second, a newly developed autoregressive network, namely the Masked Autoencoder for Distribution Estimation (MADE)~\cite{germain:2015made}, is chosen to show our experiment-to-prediction framework by taking a ensemble of micro-states for the 2-D Ising model from Metropolis simulation at a given temperature as a set of experimental measurements under a certain condition. Surprisingly, it will be shown that the machine-learned Hamiltonian encoded in neural networks would predict phases at different temperature correctly with very small number of configurations. 
	Finally we generalize the idea of the treatment to coarse-grained {\it d.o.f.} corresponding to experimental ones. As the third law of progress in theoretical physics presents\cite{weinberg:1983why}, \textit{"one could choose any degree of freedom to model a physical system, but if a wrong one used one would be sorry."} However the situation is not so bad for a machine. We will show that an alternative \textit{d.o.f}, which may theoretically vague but closer to experiments, would also work reasonably good if measurements are fine enough.
	
	\begin{figure}[!hbt]
		\centering
		\includegraphics[width=9cm]{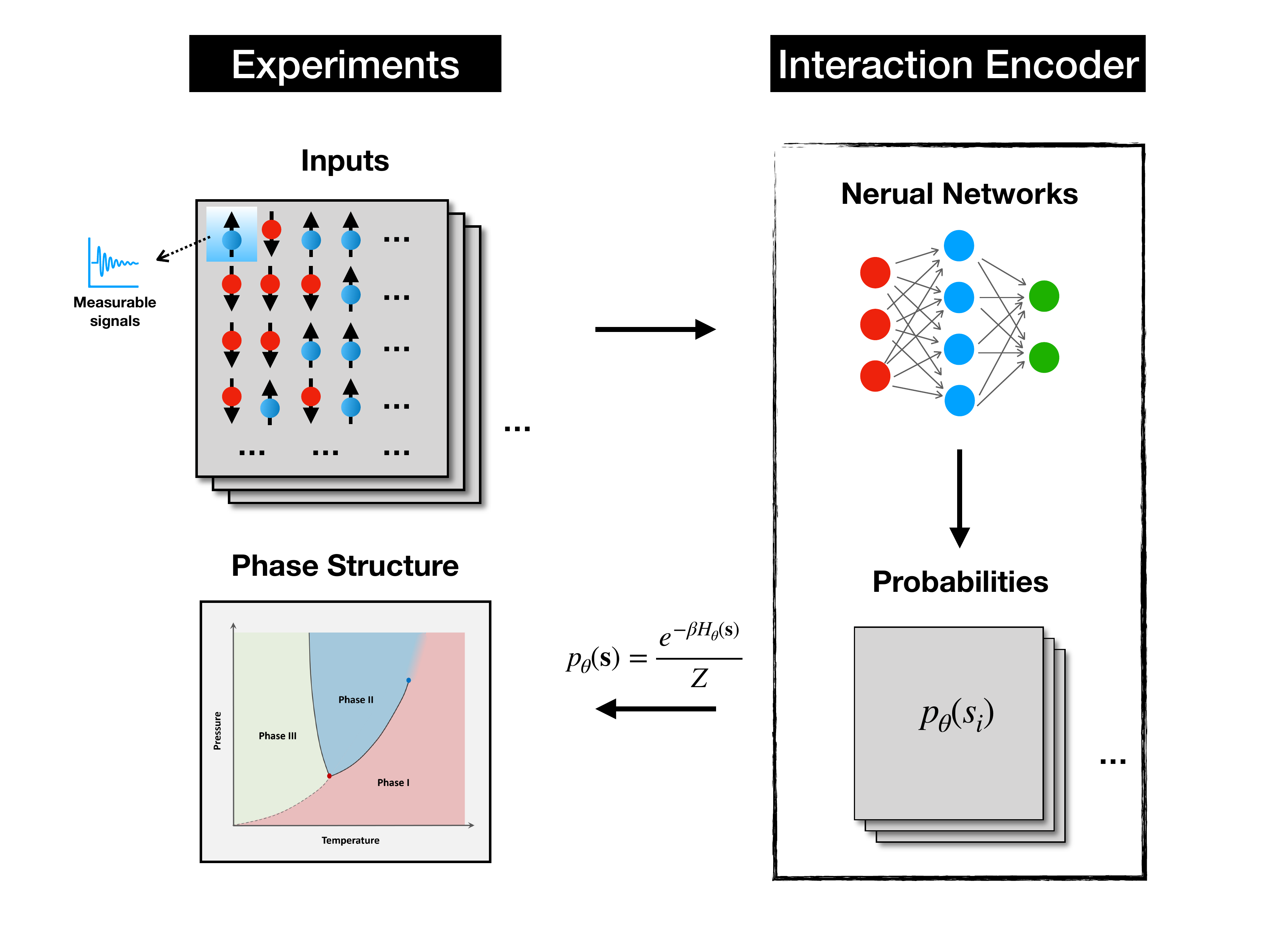}
		\caption{A paradigm of the Neural Network Statistical Mechanics. The inputs are collected from experiments or the first-principle calculations. As for the 2-D spin systems, the inputs are magnetic domain fed to the encoder. The interactions are encoded by the neural networks which produces the probability distributions of the microscopic states. The configurations are sampled at different external parameters with the well-trained networks, which predicts the phase structure.}
		\label{encoder}
	\end{figure}
	
	{\it Neural network statistical mechanics.---}
	The ensemble theory claims that a macroscopic state corresponds to a set of microscopic states which distribute according to the Boltzmann factor $e^{-H/T}$, where $H$ is the energy of the micro-state and temperature $T$ of the system. Obviously the term micro-state indicates that the system have to be further deconstructed into certain kind of \textit{d.o.f} whose choice is usually not unique. Taking a sample of magnetic material for example, the potential \textit{d.o.f} would be the local circular current, the magnetic moment of artificial divisions, the magnetic moments of electron and nuclei or the simplest Ising spin. Naturally different choices will lead to different Hamiltonian/interaction for them, because the macroscopic observable are definitive. This means combining a suitable \textit{d.o.f} ${s_i, i=1, ..., N}$ with an elaborate Hamiltonian $H(s_1, s_2, ..., s_N)$ would be enough to compute any macroscopic quantities with the help of a sound sampler for the joint distribution $p(s_1, s_2, ..., s_N)\propto exp[-H(s_1, s_2, ..., s_N)/T]$. Because of the \textit{d.o.f} choice and the corresponding Hamiltonian working in a complementary way in the framework, in the traditional approach the \textit{d.o.f} have to be chosen very carefully to avoid too complicated interactions. 
	
	Now if we ignore both the aesthetic pursuit and the limit of computing capability, it could be noticed that there are only two points are necessary for thermodynamic properties, i.e. the \textit{d.o.f} labeling different micro-states and the Hamiltonian/energy for each state. Once they are achieved, even neither the most economic nor elegant, macroscopic observables as functions of environment parameters, such as temperature and chemical potential, would be able to be computed with the distribution/Boltzmann factor of micro-states in several sound ways, such as the well-known Metropolis simulation. As the energy/Hamiltonian of each micro-state is supposed to be coded in the ensemble at any one temperature because of the Ergodic hypothesis, it is possible for the neural network to learn the distribution from the ensemble of a system and thus give the Hamiltonian of each micro-state no matter which \textit{d.o.f} adopted to deconstruct the system. Ideally our paradigm is a experiment-to-prediction one if the input ensemble could be obtained directly from measurements. And as an anticipatable byproduct the experimental noise and fluctuations would be taken good care of by this approach because of the inherent robustness of the neural network to them.

	In Fig.\ref{encoder}, a full flow-chart is proposed for describing the scheme of the Neural Network Statistical Mechanics. The left part of the sketch is the experimental port, in which the configurations are collected to feed the following machine. The so-called configurations could be measurable signals in experiments, such as the measurable signals detected by the TEM/SEM/SPM~\cite{ge:2020deep}, or the configurations sampled from a first-principle computation algorithm, such as the Markov Chain Monte-Carlo(MCMC) on lattice. For the sake of simplicity, the 2-D spin system is chosen to test the new mechanism, where the inputs are the micro-configurations generated by the MCMC. With respect to the right part, the interaction encoder is constructed by the neural networks to the computation port. The networks are arbitrarily chosen in principle, in which the representative ability is the first-line consideration. The outputs of the encoder are the probability for each configuration in the whole ensemble, which is actually an estimation from the sample. To train the machine is to reduce the loss function built in reaching the real distribution of input configurations. In the 2-D spin system case, it is derived from the cross-entropy, the loss function $ \mathcal{L} = -\sum_{\mathbf{s}\sim q_{data}}\log(p_\theta(\mathbf{s}))$, where the $\mathbf{s}^{(j)} = \{s_1^{(j)},s_2,\dots,s_N^{(j)} \}$ is the spin orientation on lattice from the training data set $j$ batch with distribution $q_{data}$, and $p_\theta(\mathbf{s})$ is the likelihood of the configuration $\mathbf{s}$ with parameters of the neural network $\{\theta\}$. The well-trained encoder is equivalent to the Hamiltonian. The last step is to generate configurations by an arbitrary algorithm with the Interaction Encoder help at different external parameters. The end-to-end machine in Fig.\ref{encoder} can predict the phase structure with only the Boltzmann distribution as a prior knowledge. Considering the interaction emerges with non-linear active functions in the neural networks (see 
	Appendix A.), it is a natural constraint to choose an autoregressive structure. In our case, the MADE is used as a distribution estimator to extract the interaction from raw configurations. The MADE is an highly efficient distribution estimator \cite{germain:2015made}, which is widely applied in several classification projects especially in the image recognition as the other autoregressive models did\cite{salimans:2017pixelcnn,ou:2019review}. 
	
	\begin{figure}[!hbt]
		\centering
		\includegraphics[width=6cm]{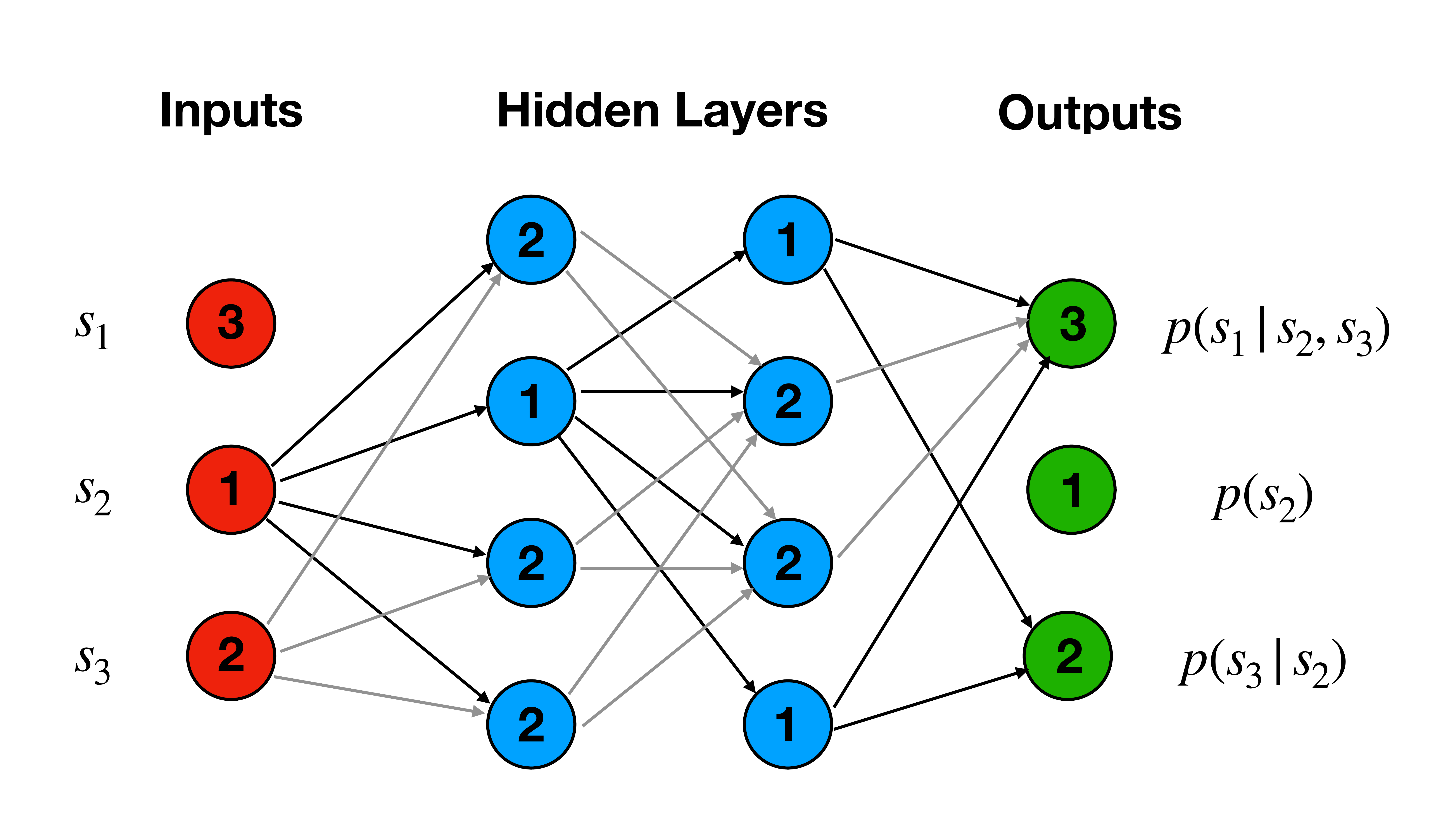}
		\caption{Deep MADE. As a concrete example, a 4 layers MADE network is presented, in which input layer contains 3 nodes with the micro-state of configurations. The number labels the order based on the conditional probability, however it is not necessary to be set as same as the site order. The masks constructed by order ensure that MADE satisfies the autoregressive property, allowing it to form a meaningful probability, here it is $p(\mathbf{s})=p(s_1|s_2,s_3)p(s_3|s_2)p(s_2)$. Connections in dark black correspond to paths that depend only on 1 input, while the light black connections depend on 2 inputs.}
		\label{made}
	\end{figure}

	{\it Interaction encoder.---}
	The Interaction encoder is established with the MADE model shows in Fig.\ref{made}. The structure of the network is the same as a generic autoencoder\cite{germain:2015made,ou:2019review}, in which a set of connections is removed such that each input unit is only decided from the previous ones by using multiplicative binary masks. The following discussion is based on 2-D spin system but can be easily extended to the other physics systems as the previous section point out. As a typical machine learning project, the data set is generated from a classical Monte-Carlo algorithm with 60000 configurations and divided into 128 batches in 2-D Ising model. In the following calculations, the default setup of the network we adopted in MADE is with input(and output) nodes as $N=(16\times16)=256$ and with two hidden layers $(180,120)$. 
	To train the Interaction encoder is to reduce the loss function
	\begin{equation}\label{loss}
	\mathcal{L}=-\sum_{\mathbf{s}\sim q_{data}}\sum_{d=1}^N \log(q_{\theta}(s_{d} | \mathbf{s}_{<d}))
	\end{equation}
	where the likelihood for each configuration is represented by the combination of the conditional probability  $p_\theta(\mathbf{s})=\prod_{d=1}^{N} q_{\theta}\left(s_{d} | s_{1}, \ldots, s_{d-1}\right)$, and $d$ labels the order of nodes in the output layers as Fig.\ref{made} shows the number in circle. As the outputs of the MADE, $p_\theta(\mathbf{s})$ is a relative accurate estimation to the real data distribution $exp[-H(\mathbf{s})/T]$(as Ref.~\cite{lin:2017why} mentioned, but autoregressive networks give a more rigorous definition.). Up to a normalization constant, the corresponding MADE Hamiltonian is
	\begin{equation}\label{ham}
	H_\theta(\mathbf{s})=-T\log p_\theta(\mathbf{s})
	\end{equation}
	
	\begin{figure}[!hbt]
		\centering
		\includegraphics[width=7cm]{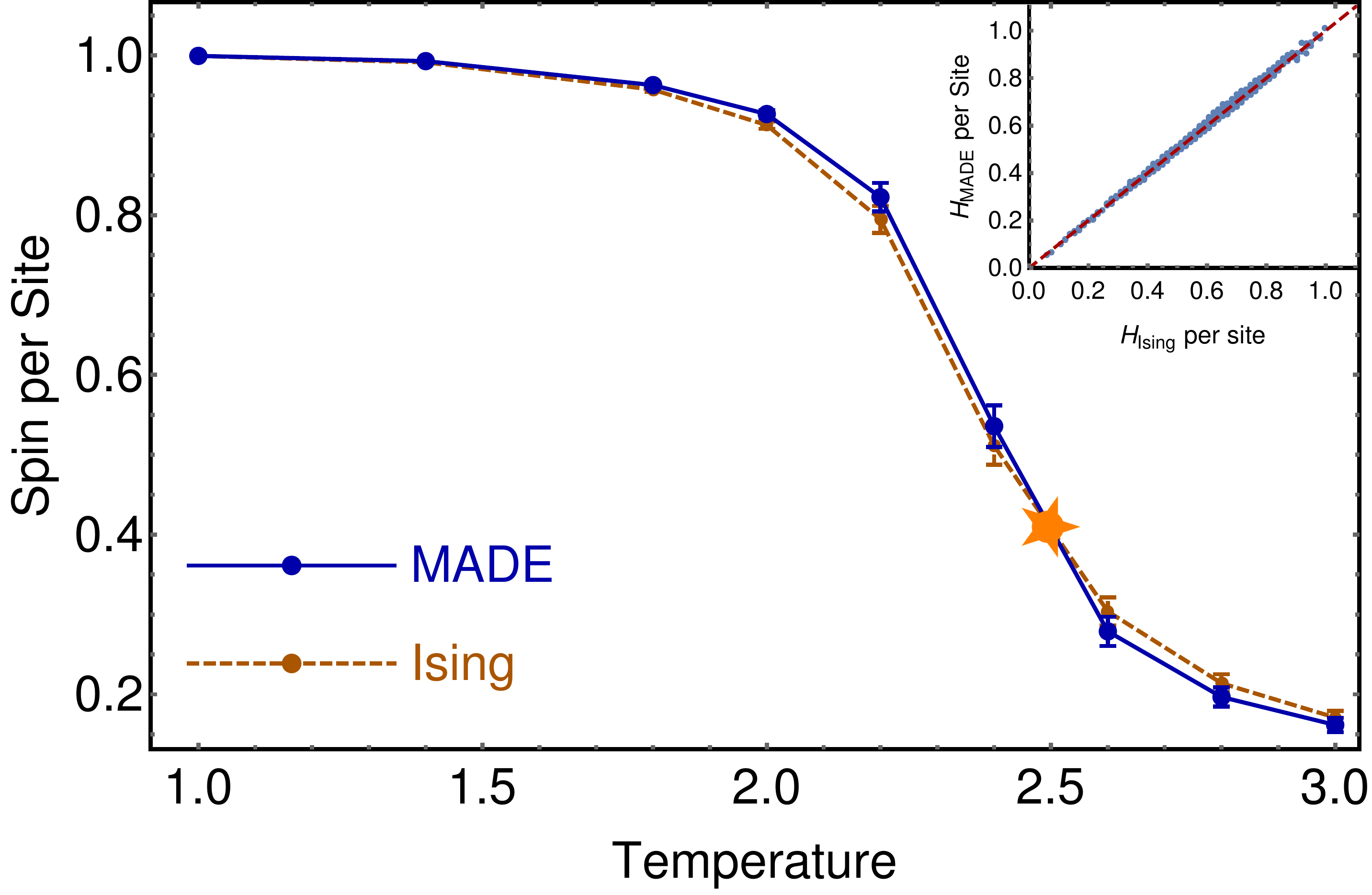}
		\caption{The magnetization density $M/V$ or the spin per site as a function of temperature obtained by the Hamiltonian of 2-D Ising model(orange dashed line) $H(\mathbf{s})=-\sum_{<i,j>}s_is_j$ and the MADE Hamiltonian(blue solid line) $H_\theta(\mathbf{s})$. And the comparison of the two Hamiltonians(top right corner).}
		\label{f2}
	\end{figure}
	
	In Fig.\ref{f2}, the MADE Hamiltonian is extracted at $T=2.5$ shown as the orange star, and the energy distribution is shown in the sub-figure, in which the unsupervisedly modeled MADE Hamiltonian matches well the underlying Ising system energy for each configurations in the ensemble (we checked that for Ising spin systems go beyond two-body spin interactions our method can also give an automatic reasonably good interaction encoder). The other blue points are generated by MCMC with the MADE Hamiltonian, in opposite, the dark orange points are all generated by MCMC with 2-D Ising model. From the narrow error bar and the behavior near the phase transition point, the MADE Hamiltonian achieves the same ability of expression as the Ising model for this 2-D spin system.
	
	{\it Effective degree of freedom detection.---}
	Using the 2-D spin system we have already shown that the MADE could fit the Hamiltonian well by making use of a physical ensemble of micro-states, and give correct predictions in a wide range of temperature. If the ensemble is treated as experimental measurements in our paradigm, a natural question is what about the measurement which is done by a device with lower resolution or whether the choice of \textit{d.o.f} as the fundamental one, such as the Ising spin here, is necessary for thermodynamics. In principle one could describe a system with many possible \textit{d.o.f}. Analytically a different choice would result in a too complicated interaction, such as the Van der Waals potential to the QED\cite{buhmann:2013dispersion} and nucleon force to QCD\cite{ishii:2007nuclear}. 
	
	In order to show the dependence of thermodynamics on the \textit{d.o.f} choice and the predictive capability of our paradigm, a lower resolution ensemble for input is generated by implementing a block transformation to each configuration, that is taking every $2\times 2$ block of $s_{ij}$ as the effective \textit{d.o.f} $S_{IJ}$. Thus all the configurations $\{(s^{(n)}_{ij})_{1\leq i, j\leq 16}\}$ are transformed into $\{(S^{(n)}_{IJ})_{1\leq I, J\leq M}\}$, where $n$ is the index of configuration and $M$ depends on the stride which is the distance between spatial locations where the block summation is applied. $M=15$ if the stride is 1, while 8 if it is 2, explicitly 
	$S_{I J}=s_{i, j}+s_{i+1, j}+s_{i, j+1}+s_{i+1, j+1}$
	where $i=1+d(I-1)$, $j=1+d(J-1)$ and $d$ is the value of stride. Obviously $S_{IJ}$ could take values in $\pm 4$, $\pm 2$ and $0$ instead of $\pm 1$. Actually, this transformation, which is known as the Kadanoff transformation as well, can also be done by neural networks\cite{mehta:2014exact}.
	
	\begin{figure}[!hbt]
		\centering
		\includegraphics[width=7cm]{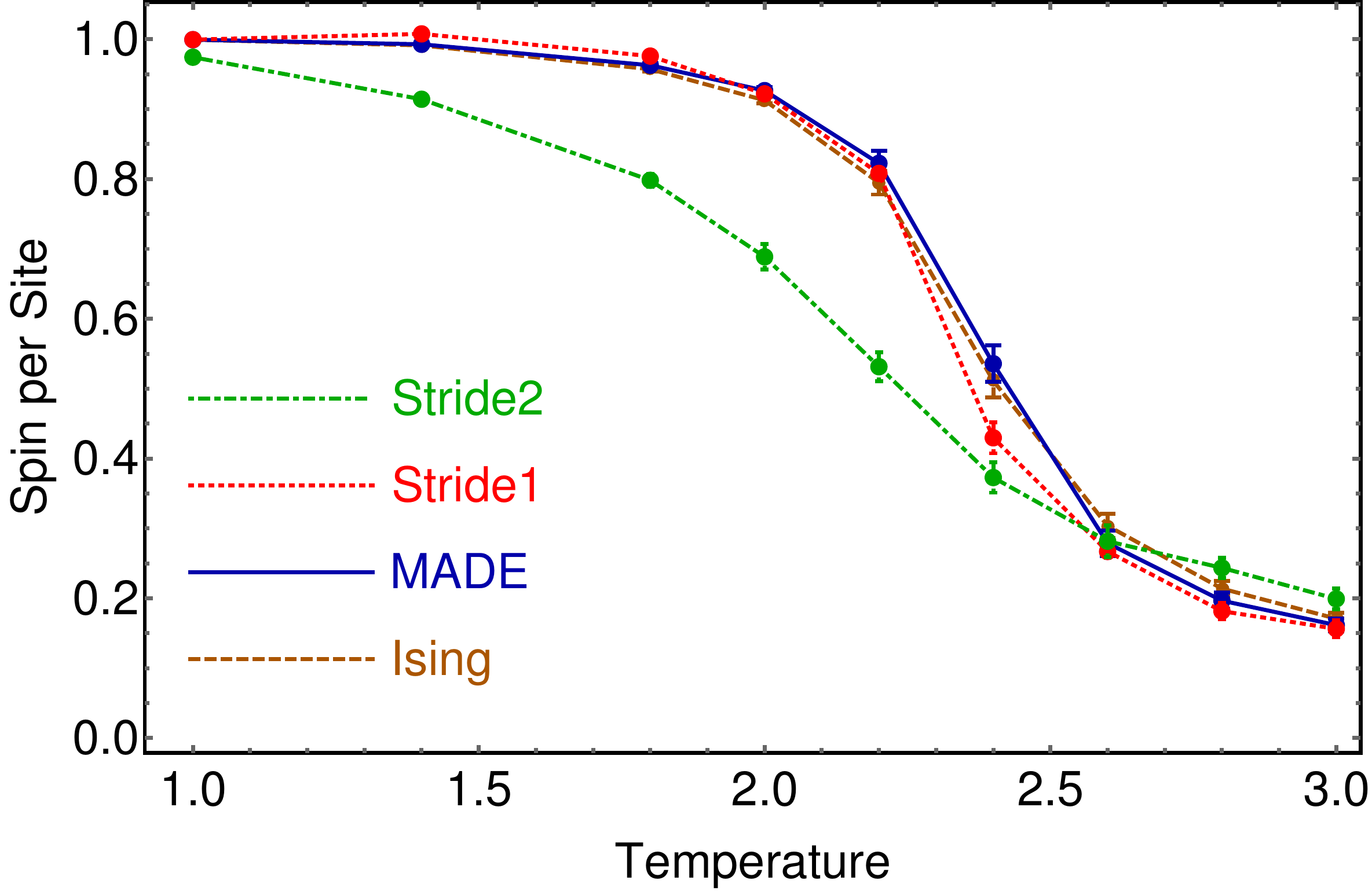}
		\caption{The spin per site as a function of temperature obtained by the standard Hamiltonian of Ising model(Dashed), the MADE Hamiltonian(Solid) for the Ising spin, the MADE Hamiltonian for the coarse grained Ising spin with stride as 1(Dotted) and 2(Dot-Dashed).}
		\label{f3}
	\end{figure}
	
	The stride 2 and 1 cases are correspond to two kinds of measurement. In the former one neither the field of vision nor scanning stride is precise enough but more economic, while the later one could scan the system more precisely. After swallowed these two ensembles our machine will encode the interactions for the coarse-grained \textit{d.o.f} $S_{I J}$ in the network and thus determine the thermodynamics at different temperatures. The numerical results in Fig.\ref{f3} shows the stride 2 case could work reasonably well and the stride-1 case could give almost the same result as the original one except a small gap around the transition section. Not surprisingly, the divergence is mainly induced by the finite size of the system. Because in such a system a coarser \textit{d.o.f} will introduce more degeneracy between coarse grained configurations which will modify the underlying distribution of the ensemble. The larger the system size, the smaller the coarse-grain-induced degeneracy and the better the performance of the coarse \textit{d.o.f}. And with a smaller stride than the block size the degeneracy is lower and more details of the system have been compensated with this scheme(stride 1), therefore a better prediction has been achieved. Considering the inevitable finite-size in laboratory our framework is also a potential method to explore the existence of more fundamental \textit{d.o.f}s.
	
	{\it Conclusion and outlook.---}
	In this work, we suggested a new paradigm to thermodynamic studies by introducing a specific type of neural network for distribution estimation. It is a straightforward experiment-to-prediction framework which could learn the probability and thus Hamiltonian/energy of each microscopic state of a system which characterized by the experimental \textit{d.o.f} and provide predictions in different other environments by explicitly tuning parameters, such as temperature,  in the Boltzmann factor for each configuration. It should be mentioned that recently some works are discussing the related topics ~\cite{yau:2020generalizability,canatar:2020statistical,bachtis:2020mapping}, but our work has clearly shown and realized the framework with the autoregressive neural networks.
	
	Different from the traditional theoretical physics or \textit{Ab initio} computation, our approach is designed to be solely in the language of experimental d.o.f, such as magnetic domains, local currents, and so on, according to experimental capability and convenience instead of any abstract or fundamental \textit{d.o.f}s. which are difficult to observe directly.  With the 2-D spin system as an example, the networks have correctly established the mapping between microscopic configurations and their Hamiltonian/energy and deservedly the phase structure. And it is worth to be emphasized that this approach would become trivial if the number of configurations for training is at the same order of the complete ensemble, such as $2^{256}\sim 10^{25}$ for the $16\times 16$ 2-D spin system. In this framework, only tens of thousands micro-states are used, which means it is a highly non-trivial and efficient approach, especially for the system of continuous \textit{d.o.f} whose phase space is infinite dimensional in principle. That reminds us that this strategy could be spontaneously applied in systems whose underlying mechanism is complicated or unclear, such as searching reliable high temperature superconductivity materials, as other machine learning methods shown\cite{stanev:2018machine,dassarma:2019machine}, since precise theoretical models or numerous experiments are not necessary here. Furthermore, by implementing a block transformation to configurations of the Ising spin, we have explored the generalization ability of the framework on the choices of \textit{d.o.f}. Obviously this treatment corresponds to low-resolution experiments whose measurements are presented with some composite \textit{d.o.f}s. This work shows that the lower-resolution measurements with smaller scanning stride would produce a quantitatively accurate prediction and even the larger stride one would qualitatively reproduce the phase transition. On one hand, the difference performance between the two coarse-grain schemes suggests the finite-size issue would weaken the predictive capability with a larger-size composite \textit{d.o.f}. On the other hand, it also indicates that this approach could help one to determine the existence and size of the more fundamental and relevant \textit{d.o.f} with lower-precision devices just by scanning the sample with a stride as small as possible. 
	
	It should be mentioned that although all the procedures are established in the classical case, this paradigm could be applied to the quantum case straightforwardly by replacing the temperature dependence $H/T$ with the summation over imaginary time slides, since the dependence of the configuration trajectories on temperature is explicit in the quantum case as well\cite{liu:2019solving}. This guarantees the applicability of the two main procedures in this paradigm, i.e., learning the Hamiltonian/components with an ensemble and prediction by tuning the temperature explicitly. With regarding the path integral, input configurations could be the possible trajectories and the effective \textit{d.o.f.} can be extracted, which is also embedded into this paradigm. Another paper on the quantum case is in progress.

	~\
	\begin{acknowledgments}
		We thank Xingyu Guo and Shoushu Gong for useful discussions. The work on this research is supported by the National Natural Science Foundation of China, Grant No. 11875002(Y.J.) and No.11775123 (L.W.), by the Samson AG and the BMBF through the ErUM-Data project for funding (KZ), by the Zhuobai Program of Beihang University(Y.J.).
		
	\end{acknowledgments}
	
	\bibliography{machine_model.bib}
	\appendix
	\section{Appendix A. Constraint on the neural networks}
	
	Some constraints on the structure of the potential network could be derived with a simple example by reformulating the Boltzmann factor as multiplication of single-body conditional probabilities. Such a form is easily encoded with most of networks which are built pixel-wisely for image processing.  A 1D spin sytem with 3 sites is enough for us to show the constraints. The probability of a certain configuration ${s_1, s_2, s_3}$ is
	\begin{eqnarray}
	p(s_1, s_2, s_3)\propto exp(-\frac{s_1 s_2+s_2 s_3+s_3 s_1}{T})
	\end{eqnarray}
	where we adopt the periodic boundary condition and set the coupling $J=1$. As a trade-off between the complete joint distribution $p(s_1, s_2, s_3)$ and absolute decoupling as independent distributions $p(s_1)p(s_2)p(s_3)$, the following form could be achieved
	\begin{eqnarray}
	p(s_1, s_2, s_3)=p(s_i)p(s_i|s_j)p(s_k|s_i, s_j)
	\end{eqnarray}
	where $\{i, j, k\}$ could be any one of $\{1, 2, 3\}$ permutations.  Obviously the interaction/coupling are coded in the conditional probabilities. And the sequence of dependence are supposed to be chosen randomly, i.e. the form $p(s_2)p(s_1|s_2)p(s_1|s_2, s_3)$ should work as well as $p(s_3)p(s_1|s_3)p(s_2|s_1, s_3)$. If we choose $\{i, j, k\}=\{1, 2, 3\}$ the distribution is factorized as
	\begin{eqnarray}
	&&p(s_1)\propto e^{-2s_1-1}+2 e+e^{2s_1-1}\nonumber\\
	&&p(s_2|s_1)\propto \frac{e^{-s_1 s_2}(e^{s_1+s_2}+e^{-s_1-s_2})}{e^{-2s_1-1}+2 e+e^{2s_1-1}}\\
	&&p(s_3|s_1, s_2)\propto \frac{e^{-s_3(s_1+s_2)}}{e^{s_1+s_2}+e^{-s_1-s_2}}\nonumber
	\end{eqnarray}
	Obviously if starting from an ensemble containing $N$ configurations $\{(s^{(i)}_1, s^{(i)}_2, s^{(i)}_3), i=1, ..., N\}$ the network could have successfully learned
	$p(s_1)$ by focusing on the 1st site, $p(s_2|s_1)$ on the 2nd site by considering the state on the 1st site and $p(s_3|s_1, s_2)$ on the 3rd site by considering states on the 1st and 2nd site, the hamiltonian/energy of any configuration could be thus obtained by $H={\rm const} -T \ln(p(s_1, s_2, s_3))$, where the
	global constant corresponding to the normalization should depend on the architecture of the network and will not cause problem in further thermodynamic studies. During the reformulation there are two constraints on the network architecture. First as there are terms like $e^{-s_1 s_2}$ in the conditional probabilities, a network as $y=\sigma(L(x))$ would not work, where the $\sigma(\cdot)$ is a nonlinear layer and $L(\cdot)$ is a general linear layer, such as
	full-connecting and convolution layer. There is suppose to be at least two nonlinear layers to fit the exponent function as well as the coupling term $s_1 s_2$.
	Second as it has been mentioned that different sequences of the conditional probabilities should be equivalent practically, i.e. one should work as the same well
	as the others. In this work we have chosen the MADE as the distribution estimator. It could be seen that there are more than two nonlinear layers in this network. And the equivalence of different factorization scheme have also been checked in both this work and numerous applications in image processing.
	
\end{document}